\documentclass{optica-article}

\journal{opticajournal} 

\articletype{Research Article}

\usepackage{lineno}
\usepackage{physics}
\usepackage{amsmath}
\usepackage{fontspec}

\begin{document}

\title{Atom Optics for Multidimensional Raman Interferometry}

\author{Yingpeng Zhao\authormark{1}, Yiyang Shao\authormark{3,4}, Wenjian Tong\authormark{2}, Shuning Bao\authormark{2}, Yuanlong Zhang\authormark{2}, Mengnan Yi\authormark{2}, Shengzhe Wang\authormark{1}, Ke Shen\authormark{1}, Bin Wu\authormark{2}, Yanying Feng\authormark{1,$*$}}

\address{\authormark{1}A-Knows Lab, State Key Laboratory of Precision Measurement Technology and Instruments, Department of Precision Instruments, Tsinghua University, Beijing 100084, China.}
\address{\authormark{2}The Key Laboratory of Quantum Precision Measurement of Zhejiang Province, School of Physcis and Optical Engineering, Zhejiang University of Technology, Hangzhou, 310023, China}
\address{\authormark{3}School of Physics and Astronomy, Shanghai Jiao Tong University, Shanghai 200240, China}
\address{\authormark{4}Tsung-Dao Lee Institute, Shanghai Jiao Tong University, Shanghai 200240, China}

\email{\authormark{*}yyfeng@mail.tsinghua.edu.cn}


\begin{abstract*} 
The coherent control of atomic wave packets in multiple momentum dimensions is a central challenge in Raman atom optics and a key requirement for multidimensional atom interferometry. In this paper, a momentum-basis theoretical framework is developed for two-dimensional Raman interactions. We formulate the atom-laser interaction for two-dimensional Raman beams and obtain the effective two-level ground-state Hamiltonian after adiabatic elimination of the excited states. The resulted Hamiltonian is constructed on the reachable momentum-state lattice and includes AC Stark shifts, intra-dimensional Raman couplings, and cross-dimensional Raman coupling terms, thereby describing multidimensional Raman dynamics beyond a simple superposition of independent single-dimensional models. The framework is used to analyze the principles and operating conditions of two-dimensional atom interferometry. Cross-dimensional Raman coupling and sequential reverse Raman transitions are identified as the main mechanisms that redistribute atoms into non-target momentum states and reduce the contrast of atomic interferometer fringe. Experimentally, Ramsey fringes are observed in a cold-atom fountain interferometer, with a contrast of $14.8\%$ at an interrogation time of $T=10~{\rm ms}$, in good agreement with the theoretical calculation. The conditions required for velocity-sensitive multidimensional atom interferometry are further clarified, including detuning control, velocity-class selection, and suppression of undesired inter-dimensional transitions. This work provides a theoretical and experimental basis for multidimensional Raman atom interferometry.
\end{abstract*} 

\section{Introduction}
Advances in matter-wave interference have established a robust foundation for quantum metrology based on atomic phase evolution~\cite{borde1989atomic,kasevich1991atomic,bongs2019taking,kasevich1992measurement}. 
With the rapid progress of laser cooling and coherent atomic manipulation techniques, light-pulse atom interferometry has become a powerful tool for precision measurements and has been extensively investigated in inertial sensing and navigation~\cite{geiger2020high,jia2025closed,salducci2024quantum}, absolute gravity measurement~\cite{tino2021testing,peters2001high}, gravity gradient detection~\cite{stray2022quantum}, geophysical exploration~\cite{wu2024construction,shettell2024geophysical}, and fundamental physics experiments~\cite{fixler2007atom,prevedelli2014measuring}. 
In this approach, the measured signal derives from the phase accumulated by the atomic interferometer, while the corresponding scale factor is defined by the atom-laser interaction, enabling intrinsic absolute calibration~\cite{cronin2009optics}. 
In recent years, research in atom-interferometric inertial sensing has progressed beyond the pursuit of enhanced single-axis sensitivity toward more sophisticated schemes that enable multi-degree-of-freedom control and the decoupled estimation of multiple physical parameters~\cite{rakholia2014dual}.

To extend atom interferometry from conventional single-axis sensing toward multi-axis inertial measurements, a variety of multi-direction interferometric configurations have been developed in recent years~\cite{wu2017multiaxis,stolzenberg2025multi}. Representative examples include pyramidal magneto-optical-trap based interferometers and two-dimensional matter-wave array architectures capable of simultaneously measuring acceleration and rotation along multiple directions within a unified experimental platform. These approaches mainly realize multi-axis sensing through multidirectional interferometer geometries and independent inertial signal along different sensitive axes. However, such multi-axis interferometric configurations are conceptually different from multidimensional atom interferometry. The former mainly relies on geometrically separated interferometer axes to achieve simultaneous inertial measurements, whereas multidimensional atom interferometry originates from multidimensional atom optics, in which Raman interactions associated with different momentum-transfer directions coherently act on a unified atomic ensemble simultaneously~\cite{barrett2019multidimensional}. In this regime, multiple momentum-transfer processes jointly participate in the coherent evolution of atomic wave packets, leading to multidimensional momentum-space coupling, multiple interference pathways, and cross-coupling between different inertial signal. Under such conditions, the interferometric dynamics can no longer be accurately described as a simple superposition of independent single-axis interferometers, thereby motivating the development of unified theoretical frameworks for multidimensional coherent evolution and phase-response analysis.

Building upon the development of multidimensional atom optics, B. Barrett et al. proposed a multidimensional framework for atom optics and interferometry, establishing the basic theoretical description of coherent Raman interactions along multiple momentum-transfer directions within a unified atomic ensemble and analyzing the geometric conditions required for inertial scale-factor decoupling~\cite{barrett2019multidimensional}. Compared with conventional single-dimensional interferometers, atom-multidimensional Raman interactions involve multiple momentum-transfer pathways participating simultaneously in the coherent evolution of atomic wave packets, leading to more complex interferometric dynamics and stronger coupling between different inertial signal. As a result, multidimensional interferometric signals become more difficult to calibrate and reconstruct accurately under coupled dynamical conditions. To further explore these multidimensional coherent processes experimentally, C. Y. Chang et al. investigated the influence of the quantization axis defined by the bias magnetic field on Raman interaction in multidimension, clarifying how the relative orientation between the magnetic field and optical propagation direction affects Raman coupling efficiency and interferometric signals~\cite{chang2023multiphoton}. Subsequently, C. Y. Liu et al. demonstrated multidirectional Raman beams splitting and multidimensional Ramsey interference using cold atoms~\cite{liu2024multidimensional}. These studies established important theoretical and experimental foundations for multidimensional atom interferometry and provided guidance for the co-design of magnetic and optical field configurations in multidimensional interferometric systems. Meanwhile, multidimensional atom interferometry has also attracted increasing interest in next-generation tensor inertial sensing and full-tensor gravity gradiometry, where multidirectional coherent manipulation within a unified atomic system is considered a promising route toward compact and highly correlated multi-component gravity sensing architectures~\cite{mukherjee2019conceptualization}.

In this work, we investigate multidimensional atom interferometry from the perspective of atom optics and atom-laser interaction dynamics. Based on a full physical-process theoretical simulation platform for cold atom interferometry developed in our previous work~\cite{zhao2025construction}, we derive a comparatively complete theoretical description of multidimensional Raman transitions and momentum-state transfer processes starting from the interaction between atoms and multidirectional Raman optical fields, and establish an effective Hamiltonian framework for coherent atom-laser interactions in multidimensional interferometric configurations. The present work provides a more systematic theoretical treatment of multidimensional Raman coupling and coherent evolution in momentum space. Based on this framework, the underlying mechanisms responsible for the reduction of Raman transition efficiency in multidimensional configurations are analyzed, and the physical conditions required for realizing multidimensional atom interferometry are further clarified. Experimentally, the study is carried out using a self-developed fountain-type cold atom interferometer~\cite{zhao2024implementation,shao2025performance}, where both copropagating and counterpropagating Raman transitions are investigated, and Ramsey interference fringes are successfully observed. The experimental observations show good agreement with the theoretical analysis and simulation results. These results provide a more complete theoretical and experimental foundation for multidimensional atom interferometry and are expected to support the further development of multidimensional inertial sensing.

\section{Theory}
\begin{figure}[ht]
	\centering\includegraphics[width=12cm]{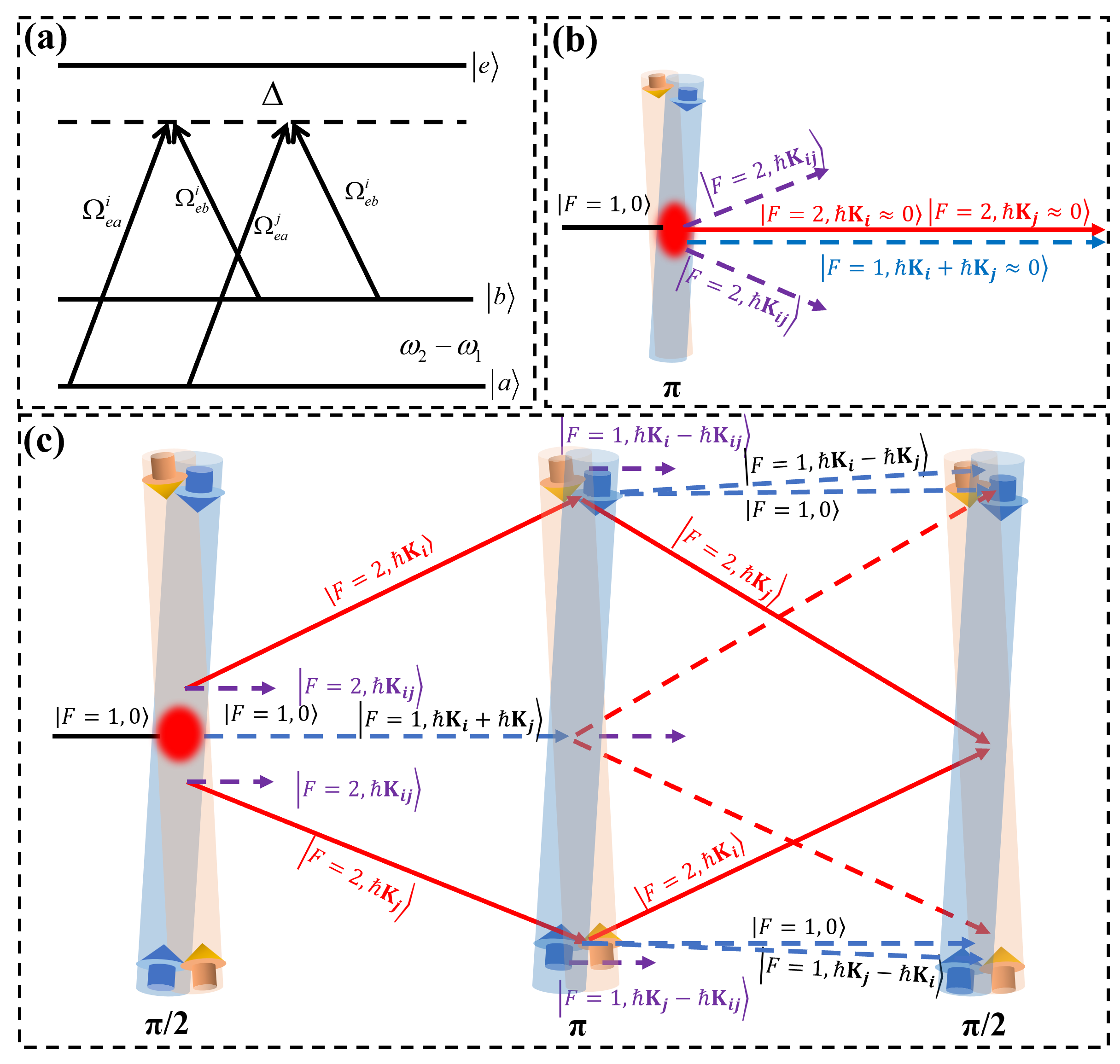}
	\caption{Schematic diagram of the physical principles: (a) Energy-level diagram of
two-dimensional Raman transitions; (b) schematic diagram of the interaction between
a two-dimensional co-propagating Raman beams and atoms; (c) schematic diagram of
the two-dimensional Mach-Zehnder interferometer.}
	\label{principles}
\end{figure}
From the perspective of atom optics, atom-laser interactions in multidimensional Raman beams can be understood through the effective Raman coupling between different momentum states. Figure 1 shows the basic principles of atom-multidimensional Raman interactions and interferometry. Compared with conventional single-dimensional Raman transitions, the introduction of additional Raman coupling directions increases the number of accessible momentum-transfer channels (As shown in Fig.~\ref{principles} (a)). In counterpropagating Raman configurations, different Raman beams directions can drive atoms into multiple momentum states through different coupling pathways. For multidimensional Mach-Zehnder (M-Z) interferometers, this leads to the formation of additional parasitic interference loops, as shown in Fig.~\ref{principles}, while part of the atomic population is transferred into momentum states that do not contribute to the target interference signal, resulting in a reduction of interferometric contrast.

In the two-dimensional M-Z interferometer shown in Fig. 1(b), the red loop corresponds to the target multidimensional interference pathway proposed in Ref.~\cite{barrett2019multidimensional}. However, the interference-path analysis shows that this closed loop mainly involves atoms remaining in the $\left| F=2 \right\rangle $ manifold during the Raman interaction process, making conventional state-selective detection schemes difficult to apply directly. In addition, after multiple off-target momentum-state transfers, the atomic population remaining in the desired interference pathway becomes significantly reduced. Nevertheless, multidimensional M-Z interferometry is still feasible. Besides the target multidimensional loop, Fig. 1(b) also contains two independent single-dimensional interference loops associated with each Raman coupling direction, which can still serve as the dominant interference pathways for multidimensional inertial measurements.

To further analyze the multidimensional Raman interaction process, we also consider the copropagating Raman transition configuration shown in Fig. 1(b). In this case, the number of involved momentum states is greatly reduced, making the multidimensional Raman transition dynamics easier to analyze. The blue dashed lines in Fig. 1(b) indicate that Raman coupling along different directions can drive part of the transferred atomic population back toward the initial momentum states through reverse transition processes. This reverse population transfer is one of the main reasons for the reduction of Raman transition efficiency in atom-multidimensional Raman interactions.

Next, we derive the optical-field Hamiltonian for two-dimensional atom interferometry and analyze the coherent evolution of atoms under atom-multidimensional Raman interactions. The detailed derivation is provided in the supplemental document.

In the two-dimensional Raman configuration considered here, the indices $i$ and $j$ label two independent Raman directions. Along each direction, two optical frequency components are used to drive the Raman transition between the two ground states $|a\rangle$ and $|b\rangle$. We denote the single-photon wave vectors of these components by $\boldsymbol{\kappa}_{ia}$ and $\boldsymbol{\kappa}_{ib}$ for direction $i$, and by $\boldsymbol{\kappa}_{ja}$ and $\boldsymbol{\kappa}_{jb}$ for direction $j$. The intended Raman momentum transfers are therefore

\begin{equation}
\mathbf k_i=\boldsymbol{\kappa}_{ia}-\boldsymbol{\kappa}_{ib},
\qquad
\mathbf k_j=\boldsymbol{\kappa}_{ja}-\boldsymbol{\kappa}_{jb},
\end{equation}
Here $\mathbf k_i$ and $\mathbf k_j$ are the effective two-photon Raman wave vectors for the two designed dimensions. Since the Raman fields associated with different directions are applied simultaneously, optical frequency components belonging to different dimensions can also form cross-Raman pairs. These give rise to the additional effective wave vectors
\begin{equation}
\mathbf k_{ij}=\boldsymbol{\kappa}_{ia}-\boldsymbol{\kappa}_{jb},
\qquad
\mathbf k_{ji}=\boldsymbol{\kappa}_{ja}-\boldsymbol{\kappa}_{ib}.
\end{equation}
The first two wave vectors describe the intended two-dimensional Raman momentum transfers, whereas the latter two describe cross-dimensional momentum transfers. These cross channels may be far off resonance and negligible under suitable frequency design, but they must be included in the general Hamiltonian before applying the rotating-wave or large-detuning approximation.

The atomic basis state is written as $|\gamma,\mathbf p\rangle=|\gamma\rangle\otimes|\mathbf p\rangle$, where $|\gamma\rangle$ denotes one of the selected internal states $|a\rangle$, $|b\rangle$, or $|e\rangle$, and $|\mathbf p\rangle$ denotes the center-of-mass momentum state. Because the Raman interaction changes both the internal state and the atomic momentum, the reachable momentum states form a lattice generated by all allowed two-photon momentum transfers,
\begin{equation}
\mathcal P=
\left\{
\mathbf p_0
\pm n_i\hbar\mathbf k_i
\pm n_j\hbar\mathbf k_j
\pm n_{ij}\hbar\mathbf k_{ij}
\pm n_{ji}\hbar\mathbf k_{ji}
\;\middle|\;
n_i,n_j,n_{ij},n_{ji}\in\mathbb Z
\right\}.
\end{equation}
Here $\mathcal P$ is the set of reachable momentum states, $\mathbf p_0$ is the initial reference momentum, $n_i$, $n_j$, $n_{ij}$, and $n_{ji}$ are integer momentum-transfer orders associated with the four Raman channels, and $\hbar$ is the reduced Planck constant. In practical calculations, this momentum lattice is truncated according to the pulse order, near-resonant condition, and population cutoff.

The total Hamiltonian is written as~\cite{zhao2025construction}
\begin{equation}
H(t)=H_0+H_B+H_{\rm int}(t),
\end{equation}
where $H_0$ is the free atomic Hamiltonian, $H_B$ is the Zeeman Hamiltonian induced by the bias magnetic field, and $H_{\rm int}(t)$ is the atom-laser interaction Hamiltonian. The free Hamiltonian is
\begin{equation}
\hat H_0
=
\frac{\hat{\mathbf p}^{\,2}}{2m}
+
E_a|a\rangle\langle a|
+
E_b|b\rangle\langle b|
+
E_e|e\rangle\langle e|.
\label{eq:bare_atom_hamiltonian}
\end{equation}
Here $\hat{\mathbf p}$ is the center-of-mass momentum operator, $m$ is the atomic mass, $E_a$, $E_b$, and $E_e$ are the internal energies of $|a\rangle$, $|b\rangle$, and $|e\rangle$, respectively. The tensor product indicates that the kinetic-energy term acts on the external momentum degree of freedom while leaving the internal state unchanged. The Zeeman interaction can be written as
\begin{equation}
H_B=-\hat{\boldsymbol{\mu}}\cdot\mathbf B,
\end{equation}
where $\hat{\boldsymbol{\mu}}$ is the magnetic-dipole operator and $\mathbf B$ is the bias magnetic field defining the quantization axis. In the selected hyperfine-Zeeman basis, this term gives the Zeeman energy shifts of the internal states.

The positive-frequency part of the field of two-dimensional Raman beams  is
\begin{equation}
\begin{aligned}
\mathbf E^{(+)}(\mathbf r,t)
=
\frac{1}{2}\Big[
&
\mathbf e_{ia}\mathcal E_{ia}(t)
e^{i(\boldsymbol{\kappa}_{ia}\cdot\mathbf r-\omega_{ia}t+\phi_{ia})}
+
\mathbf e_{ib}\mathcal E_{ib}(t)
e^{i(\boldsymbol{\kappa}_{ib}\cdot\mathbf r-\omega_{ib}t+\phi_{ib})}
\\
&
+
\mathbf e_{ja}\mathcal E_{ja}(t)
e^{i(\boldsymbol{\kappa}_{ja}\cdot\mathbf r-\omega_{ja}t+\phi_{ja})}
+
\mathbf e_{jb}\mathcal E_{jb}(t)
e^{i(\boldsymbol{\kappa}_{jb}\cdot\mathbf r-\omega_{jb}t+\phi_{jb})}
\Big].
\end{aligned}
\end{equation}
Where $\mathbf e_{ia}$, $\mathbf e_{ib}$, $\mathbf e_{ja}$, and $\mathbf e_{jb}$ are the polarization vectors; $\mathcal E_{ia}(t)$, $\mathcal E_{ib}(t)$, $\mathcal E_{ja}(t)$, and $\mathcal E_{jb}(t)$ are slowly varying electric-field amplitudes; $\omega_{ia}$, $\omega_{ib}$, $\omega_{ja}$, and $\omega_{jb}$ are optical angular frequencies; and $\phi_{ia}$, $\phi_{ib}$, $\phi_{ja}$, and $\phi_{jb}$ are optical phases. The atom-laser interaction is given by the electric-dipole coupling,
\begin{equation}
H_{\rm int}(t)=-\hat{\mathbf d}\cdot\mathbf E(\mathbf r,t),
\end{equation}
where $\hat{\mathbf d}$ is the electric-dipole operator and $\mathbf E(\mathbf r,t)=\mathbf E^{(+)}(\mathbf r,t)+\mathbf E^{(-)}(\mathbf r,t)$ is the total electric field.

After adiabatic elimination of the excited state $|e\rangle$, the atom-laser interaction in the ground-state manifold can be separated into the AC Stark term and the effective two-photon Raman coupling term,
\begin{equation}
H_{\rm int}(t)=H_{\rm AC}(t)+H_{\rm Raman}(t).
\end{equation}

The AC Stark term describes virtual single-photon excitation followed by de-excitation back to the same momentum state. In the reduced two-ground-state basis $\{|a\rangle,|b\rangle\}$, it can be written as
\begin{equation}
H_{\rm AC}(t)
=
\sum_{\mathbf p_n\in\mathcal P}
\left[
\delta^a_{\rm AC}(\mathbf p_n,t)
|a,\mathbf p_n\rangle\langle a,\mathbf p_n|
+
\delta^b_{\rm AC}(\mathbf p_n,t)
|b,\mathbf p_n\rangle\langle b,\mathbf p_n|
\right].
\end{equation}
Here $\delta^a_{\rm AC}(\mathbf p_n,t)$ and $\delta^b_{\rm AC}(\mathbf p_n,t)$ are the AC Stark shifts of $|a,\mathbf p_n\rangle$ and $|b,\mathbf p_n\rangle$, respectively. With the convention used above for the optical field and the single-photon Rabi frequency, they are given by
\begin{equation}
\delta^a_{\rm AC}(\mathbf p_n,t)
=
-\frac{\hbar}{4}
\sum_{\substack{s\in\{i, j\}}}
\frac{
|\Omega_{ea}^{(s)}(t)|^2
}{
\Delta_{a}(\mathbf  p_n)
}, \delta^b_{\rm AC}(\mathbf p_n,t)
=
-\frac{\hbar}{4}
\sum_{\substack{s\in\{i, j\}}}
\frac{
|\Omega_{eb}^{(s)}(t)|^2
}{
\Delta_b^{(s)}(\mathbf p_n)
}.
\end{equation}
Where $s\in\{i,j\}$ denotes the Raman dimension; $\Omega_{ea}^{(s)}$ and $\Omega_{eb}^{(s)}$ are the slowly varying single-photon Rabi frequencies coupling $|a\rangle$ and $|b\rangle$ to $|e\rangle$ through the optical field $r$; and $\Delta_a^{(s)}(\mathbf p_n)$ and $\Delta_b^{(s)}(\mathbf p_n)$ are the corresponding single-photon detunings including the kinetic-energy correction associated with the single-photon momentum transfer. For the single-photon coupling driven by the $ia$ component, the detuning can be written as
\begin{equation}
\Delta_a^{(i)}(\mathbf p_n)
=
\omega_{ia}
-
\frac{1}{\hbar}
\left[
E_e-E_a+
\frac{
|\mathbf p_n+\hbar\boldsymbol{\kappa}_{ia}|^2-|\mathbf p_n|^2
}{2m}
\right].
\end{equation}
The kinetic-energy term accounts for the Doppler and recoil contributions associated with the single-photon momentum transfer. In the far-detuned regime considered here, these momentum-dependent corrections are much smaller than the single-photon detuning itself. Therefore, when evaluating the single-photon detuning in the denominator of the AC Stark shift and Raman coupling, we use the momentum-independent approximation
$\Delta_a^{(i)}(\mathbf p_n)
\simeq
\Delta_a^{(i)}
=
\omega_{ia}
-
(E_e-E_a)/ \hbar$.
The single-photon detunings associated with the remaining Raman field components are approximated in the same manner. In the numerical model, the AC Stark contribution can be constructed as a matrix in the selected internal-state basis; the scalar shifts written above represent the reduced two-state limit.

The Raman term contains the off-diagonal two-photon couplings between different momentum nodes. Introducing the Raman-channel set $\mathcal C=\{i,j,ij,ji\}$,
the Hamiltonian can be written compactly as
\begin{equation}
H_{\rm Raman}(t)
=
\sum_{\mathbf q\in\mathcal P}
\sum_{c\in\mathcal C}
\left[
V_c(\mathbf q,t)
|\mathbf q+\hbar\mathbf k_c,b\rangle
\langle \mathbf q,a|
+
V_c^*(\mathbf q,t)
|\mathbf q,a\rangle
\langle \mathbf q+\hbar\mathbf k_c,b|
\right].
\end{equation}
Here $c$ labels one of the four Raman channels $i$, $j$, $ij$, and $ji$; $\mathbf k_c$ denotes the corresponding effective two-photon wave vector, with
$\mathbf k_c\in\{\mathbf k_i,\mathbf k_j,\mathbf k_{ij},\mathbf k_{ji}\}$; and $V_c(\mathbf q,t)$ is the corresponding effective Raman coupling coefficient.

For the intended Raman coupling along dimension $i$, one has
\begin{equation}
V_i(\mathbf q,t)
\simeq
-\frac{\hbar}{8}
\Omega_{ea}^{(i)*}(t)
\Omega_{eb}^{(i)}(t)
\left[
\frac{1}{\Delta_a^{(i)}}
+
\frac{1}{\Delta_b^{(i)}}
\right]
e^{-i((\omega_{ia}-\omega_{ib})t-(\phi_{ia}-\phi_{ib}))},
\end{equation}

The corresponding expression for $V_j$ is obtained by the substitution $i\rightarrow j$. 
When the $ia$ Raman field component and the $jb$ Raman field component are combined in a two-photon process, a cross-dimensional Raman coupling is obtained:
\begin{equation}
V_{ij}(\mathbf q,t)
\simeq
-\frac{\hbar}{8}
\Omega_{ea}^{(ia)*}(t)
\Omega_{eb}^{(jb)}(t)
\left[
\frac{1}{\Delta_a^{(i)}}
+
\frac{1}{\Delta_b^{(j)}}
\right]
e^{-i((\omega_{ia}-\omega_{jb})t-(\phi_{ia}-\phi_{jb}))},\label{15}
\end{equation}

Similarly, $V_{ji}$ is formed by fields $ja$ and $ib$. These cross-dimensional terms are not additional assumptions; they follow directly from the same electric-dipole interaction when all four Raman beams are applied simultaneously. If their two-photon detunings are much larger than the corresponding effective Raman Rabi frequencies, they can be neglected under the rotating-wave approximation; otherwise, they must be retained because they open additional momentum-transfer channels.

This formulation clarifies the physical origin of multidimensional momentum coupling. A designed Raman pulse along dimension $i$ can transfer $|F=a,\mathbf p\rangle \rightarrow |F=b,\mathbf p+\hbar\mathbf k_i\rangle$.
After this state is populated, a further Raman coupling along dimension $j$ can drive
$|F=b,\mathbf p+\hbar\mathbf k_i\rangle
\rightarrow
|F=a,\mathbf p+\hbar\mathbf k_i-\hbar\mathbf k_j\rangle$.
At the same time, cross-dimensional couplings can directly drive atoms into $|F=b,\mathbf p+\hbar\mathbf k_{ij}\rangle$ or $|F=b,\mathbf p+\hbar\mathbf k_{ji}\rangle$. Therefore, a two-dimensional Raman field does not generate only two independent single-dimensional momentum transfers. It produces a coupled momentum-space network containing intended, cascaded, and cross-dimensional pathways. The Hamiltonian must therefore be constructed on the complete reachable momentum lattice before any resonance or truncation approximation is applied. This is the essential difference between a multidimensional Raman model and a simple superposition of independent single-dimensional Raman transitions.

\section{Experiment method and setup}

\begin{figure}[ht]
	\centering\includegraphics[width=8cm]{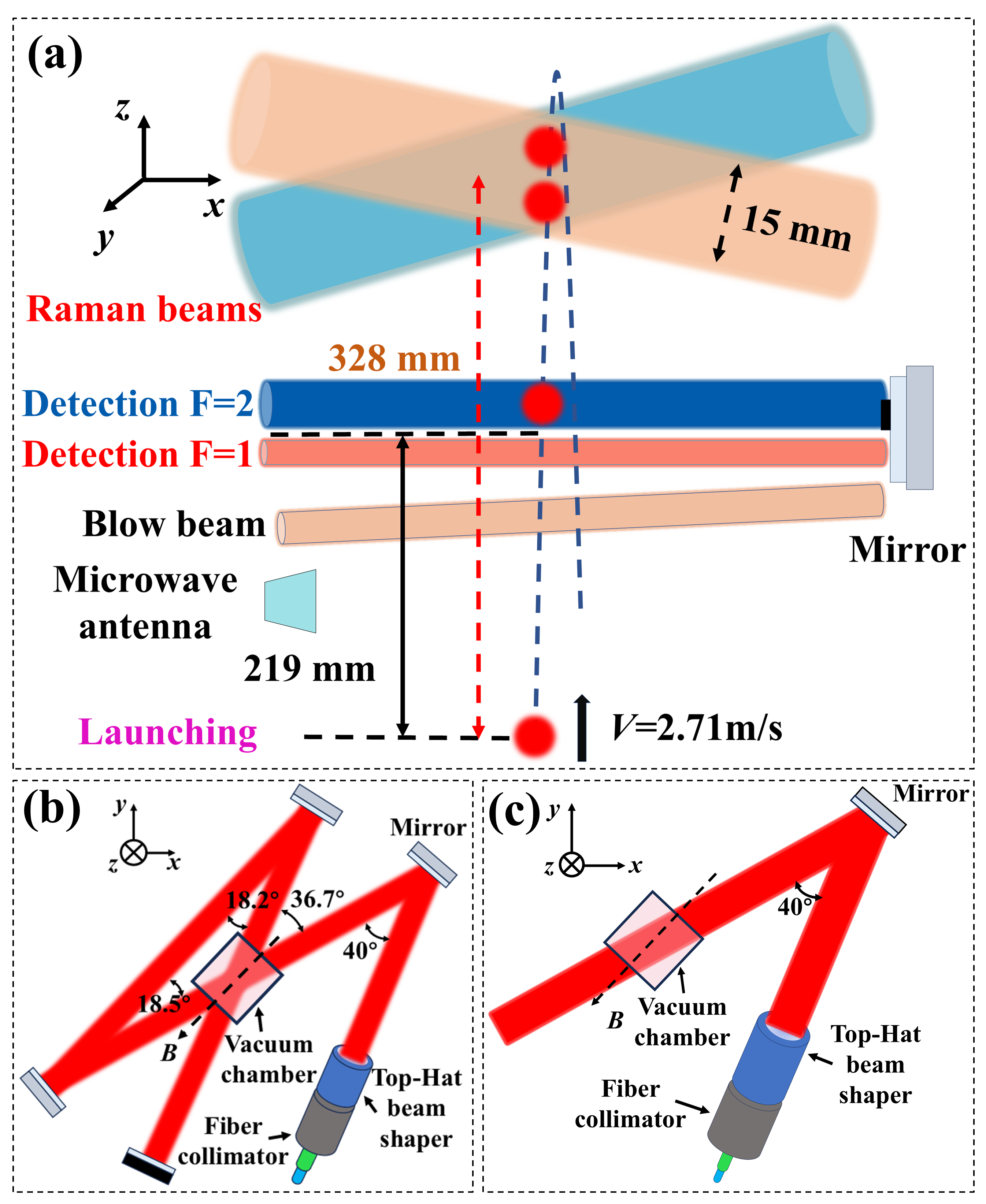}
	\caption{Two-dimensional atom interferometer experimental design. (a) Experimental sequence of the 2D Raman interferometer. (b) Structural design of the 2D Raman beams. By removing the black mirror, the optical setup can be switched to a co-propagating Raman-transition configuration. (c) Structural design of the 1D Raman beams.}
	\label{experimental apparatus}
\end{figure}

The experiment employs a quartz vacuum chamber, allowing flexible definition of the geometry of the Raman beams in both orientation and position~\cite{zhao2024implementation}. Building upon our previously developed fountain-type four-pulse cold-atom interferometer, we utilize the Raman interaction region to implement a crossed two-dimensional Raman configuration. The vacuum chamber of the interferometer is composed of a rectangular quartz-glass cell, with transverse dimensions of 55$\times$60 mm$^{2}$ in the $x-y$ plane. Due to the limited transverse size of the rectangular quartz vacuum cell, the crossing angle between the two Raman beams is set to 36.7$^\circ$. The quantizing magnetic field is aligned along the angular bisector of the two beams to ensure symmetric polarization projections. This acute-angle geometry also helps suppress undesired two-photon couplings and the associated two-photon light shifts from parasitic Raman channels. To realize the co-propagating two-dimensional Raman interferometer, the mirror highlighted in black in Fig.~\ref{experimental apparatus}(b) is removed in the experiment.

In the experiment, atoms are first cooled in a 2D MOT and transferred to a 3D MOT. Launching is achieved by tuning the cooling laser detuning, with polarization gradient cooling applied before and after the launch to reach a temperature of approximately $4~\upmu\mathrm{K}$. The atomic population is prepared in the magnetically insensitive $\ket{F=1, m_F=0}$ state using a microwave pulse, while a push beam removes residual atoms in other states. The atomic cloud is launched upward at $2.71~\mathrm{m/s}$ and reaches the Raman interaction region, located $328~\mathrm{mm}$ above the 3D-MOT center, after approximately $179~\mathrm{ms}$. Each Raman beams consist of a collimation lens and a Gaussian-to-flat-top beam-shaping optic. The 1/$e^2$ beam diameter of the collimated Raman beams is $10~\mathrm{mm}$, followed by a $15~\mathrm{mm}$ Gaussian-to-flat-top beam to ensure uniform intensity across the atomic cloud. The Raman beams are aligned at an angle of approximately $8^\circ$ relative to the horizontal (or equivalently $82^\circ$ relative to the vertical gravity direction), so that the component of the atomic velocity along the beam direction introduces a controlled Doppler shift in the counter-propagating measurement. After the Raman interaction, atoms fall to a detection region $219~\mathrm{mm}$ above the 3D-MOT center, reaching it at approximately $454~\mathrm{ms}$, and the final population distribution is read out using a dual-state fluorescence detection scheme that simultaneously measures the $\ket{F=1}$ and $\ket{F=2}$ populations.

\section{Experiment and simulation results}

\subsection{Experiment results}

\begin{figure}[ht]
	\centering\includegraphics[width=10cm]{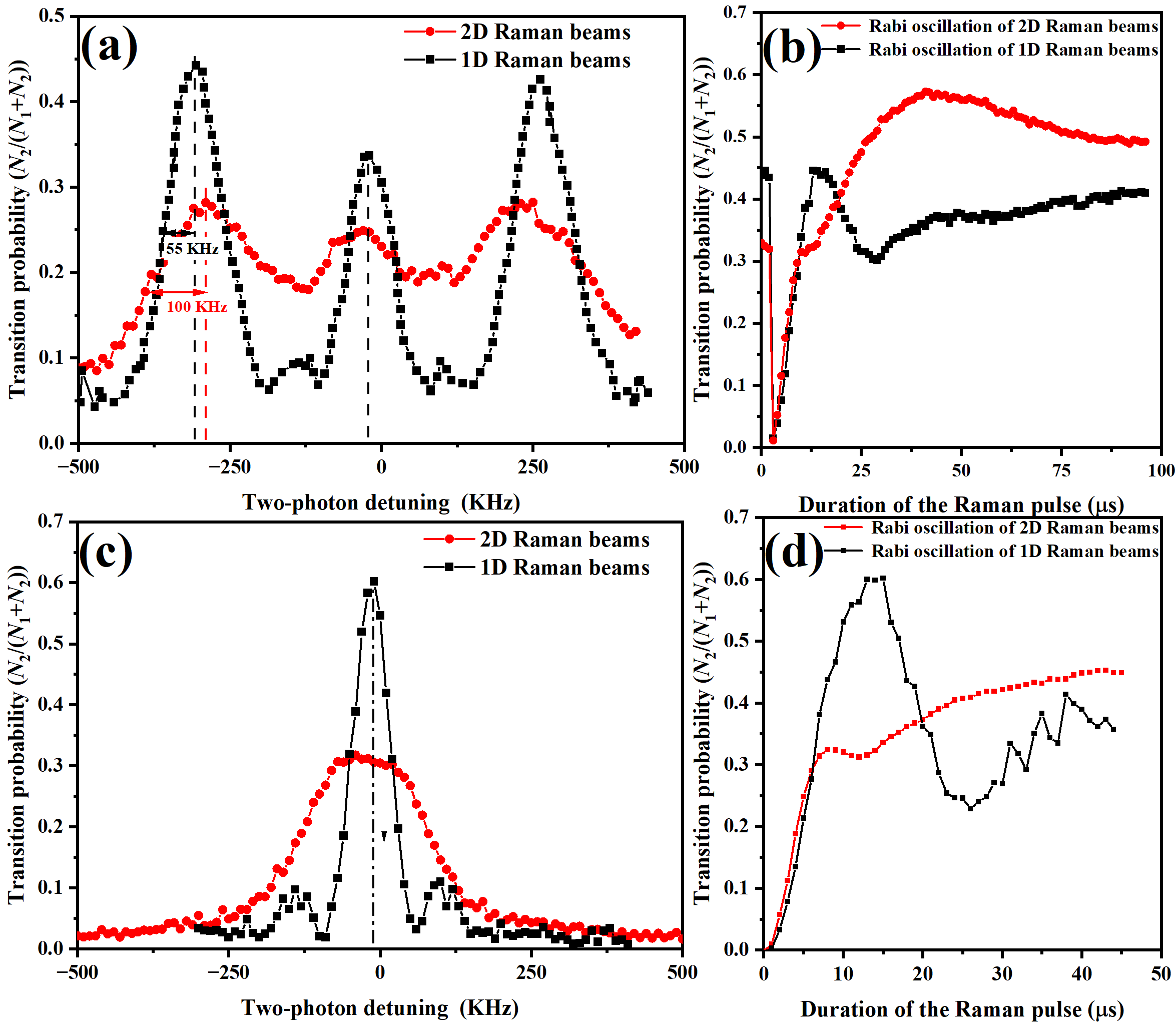}
	\caption{Experimental comparison of one-dimensional and two-dimensional Raman processes. (a) co-propagating Raman spectrum; (b) co-propagating Raman Rabi oscillation; (c) counter-propagating Raman spectrum; (d) counter-propagating Raman Rabi oscillation. The black dotted line represents the interaction between the atoms and the single-dimensional Raman beams, while the red dotted line represents the interaction between the atoms and the two-dimensional Raman beams.}
	\label{Raman spectrum and Raman Rabi}
\end{figure}
In the experiment, we first investigated the velocity-sensitive Raman transition. The Raman beam power after the optical fiber was approximately $90~\mathrm{mW}$. After passing through the Gaussian-to-flat-top beam-shaping optics, the optical power was partially reduced due to optical losses. Fig.~\ref{Raman spectrum and Raman Rabi}(a) presents the measured velocity-sensitive counter-propagating Raman spectra for the single-dimensional and two-dimensional configurations. The red dotted curve represents the two-dimensional Raman result, while the black dotted curve corresponds to the single-dimensional Raman configuration. Because the Raman beams polarizations were not fully optimized, the co-propagating resonance peaks were not completely suppressed. This allows the co-propagating and counter-propagating resonances to be compared in the same measurement. As shown in Fig.~\ref{Raman spectrum and Raman Rabi}(a), both the co-propagating and counter-propagating Raman peaks in the two-dimensional configuration are broader than those in the single-dimensional configuration. This indicates that the multidimensional Raman interaction broadens the Raman transition spectrum and reduces the Raman transition efficiency.  In addition, comparison of the co-propagating resonance peaks reveals a noticeable frequency shift. This shift is likely related to light-field-induced effects, such as differential AC Stark shifts and cross-dimensional Raman coupling, while residual magnetic-field-dependent shifts may also contribute.

The measured half width at half maximum (HWHM) of the velocity-sensitive Raman spectra is approximately $55~\mathrm{kHz}$ for the single-dimensional configuration and $100~\mathrm{kHz}$ for the two-dimensional configuration, corresponding to full widths at half maximum (FWHM) of approximately $110~\mathrm{kHz}$ and $200~\mathrm{kHz}$, respectively. The linewidth ratio is therefore about $1.82$. Part of this linewidth increase can be attributed to power broadening caused by the enhanced effective Raman coupling in the two-dimensional configuration. This is consistent with the Rabi-oscillation measurements shown in Fig.~\ref{Raman spectrum and Raman Rabi}(b). In the single-dimensional case, the first $\pi$ pulse occurs at approximately $13~\upmu\mathrm{s}$. In the two-dimensional case, the transition probability shows a characteristic response time of about $11~\upmu\mathrm{s}$, although no clear $\pi$ pulse or subsequent Rabi oscillation is observed. Using these time scales to estimate the effective Rabi-frequency ratio gives $\Omega_{\mathrm{2D}}/\Omega_{\mathrm{1D}}\approx 13/11\approx1.18$, which corresponds to an estimated FWHM of about $130~\mathrm{kHz}$. Even if the effective Raman optical power is assumed to increase by a factor of two, the expected linewidth scales only as $\sqrt{2}$, giving an estimated FWHM of about $155~\mathrm{kHz}$. Both estimates are smaller than the measured value of approximately $200~\mathrm{kHz}$, indicating that power broadening is not the only broadening mechanism. Additional contributions may come from cross-dimensional Raman couplings with different effective wave vectors, population transfer to undesired momentum states, spatially inhomogeneous differential AC Stark shifts, and intensity imbalance between the two Raman dimensions. The changes in the co-propagating Raman spectra, which are less sensitive to first-order Doppler broadening, also indicate the presence of differential AC Stark shifts. These results show that the linewidth increase in the two-dimensional Raman configuration results from the combined effects of power broadening, inter-dimensional Raman coupling, and light-shift inhomogeneity.

Fig.~\ref{Raman spectrum and Raman Rabi}(c) shows the co-propagating Raman spectra measured in the single-dimensional and two-dimensional configurations. The black trace corresponds to the single-dimensional Raman transition, where two secondary peaks located at approximately $\pm 140~\mathrm{kHz}$ are consistent with transitions involving neighboring Zeeman sublevels under a bias magnetic field of about $0.2~\mathrm{G}$. In the two-dimensional case, the Raman spectrum is significantly broadened. This broadening is mainly associated with the multidimensional Raman coupling and the additional cross-dimensional Raman channels formed by the superposition of Raman fields from different dimensions. These channels can introduce nonzero effective wave vectors, such as $\mathbf{k}_{\mathrm{eff}}=\mathbf{k}_i-\mathbf{k}_j$, and therefore lead to additional Doppler-sensitive broadening for atoms with finite velocity distributions. Meanwhile, the simultaneous application of Raman fields from two dimensions can also enhance the effective power broadening. Channel-dependent or spatially inhomogeneous AC Stark shifts may further contribute to the observed spectral broadening and frequency shift. As a result, the full width at half maximum (FWHM) increases from approximately $0.055~\mathrm{MHz}$ in the single-dimensional case to approximately $0.191~\mathrm{MHz}$ in the two-dimensional case.

In the multidimensional Raman spectrum, in addition to the AC Stark shifts determined by the intensity ratio within each individual Raman dimension, extra light shifts can arise from cross-dimensional coupling. Specifically, mixed Raman pathways formed by the $\ket{F=1}$ Raman field of one dimension and the $\ket{F=2}$ Raman field of another dimension introduce additional channel-dependent AC Stark shifts. In an ideal symmetric configuration with equal optical intensities in the two dimensions, these shifts are expected to largely cancel; however, any intensity imbalance between the two dimensions leads to residual light shifts, which can appear as both frequency shifts and spectral broadening. This effect is particularly relevant in the present folded-reflection configuration, where the Raman beams for the second dimension are generated through reflection at an acute incident angle, while the mirror coating is optimized for near-normal incidence. The additional reflections were experimentally measured to cause an optical power loss of approximately $29\%$, introducing an intensity asymmetry between the two Raman dimensions. Moreover, different momentum-state couplings in the multidimensional Raman field are associated with different effective wave vectors, so that cross-dimensional Raman channels can introduce separated Doppler shifts between different momentum states even in the co-propagating spectrum. These shifted resonance conditions can overlap with the main Raman resonance and further contribute to the observed spectral broadening.As shown in Fig.~\ref{Raman spectrum and Raman Rabi}(c), the two-dimensional co-propagating Raman spectrum exhibits a broadened and flattened central resonance compared with the single-dimensional case. The central resonance is flattened over a frequency range of approximately $70~\mathrm{kHz}$, rather than forming a sharp peak as in the single-dimensional spectrum. This plateau-like feature suggests that several closely spaced resonance contributions overlap near the line center.

Fig.~\ref{Raman spectrum and Raman Rabi}(d) shows the velocity-sensitive Rabi oscillations driven by the counter-propagating Raman beams. In the single-dimensional configuration, a clear Rabi oscillation is observed, with the first $\pi$-pulse point appearing at approximately $14~\upmu\mathrm{s}$. In the two-dimensional configuration, the first maximum appears at approximately $8~\upmu\mathrm{s}$, but the corresponding transition probability is only about $53\%$ of that obtained at the single-dimensional $\pi$-pulse point. This indicates that the multidimensional Raman field does not improve the effective population transfer efficiency in a simple two-state picture. Similar to the behavior observed in the counter-propagating Rabi measurement, part of the atomic population may be redistributed through reverse Raman transitions and cross-dimensional coupling channels, which suppress the effective $\ket{F=1}\leftrightarrow \ket{F=2}$ transition and reduce the measured Raman transition efficiency.

After characterizing the Raman spectrum and Rabi oscillations, a $\pi/2-\pi/2$ pulse sequence with an interrogation time $T$ is applied within the overlap region of the two Raman beams, thereby realizing a two-dimensional Ramsey interferometer. By scanning the two-photon detuning of the Raman transition, the corresponding two-dimensional Ramsey fringes are obtained.

To clearly distinguish the two-dimensional configuration from its single-dimensional counterpart, only the first retro-reflecting mirror is retained while the remaining mirrors are removed, thereby realizing a single-dimensional Ramsey interferometer. 
The corresponding experimental setup is shown in Fig~\ref{experimental apparatus}(c).

\begin{figure}[!hb]
	\centering\includegraphics[width=8cm]{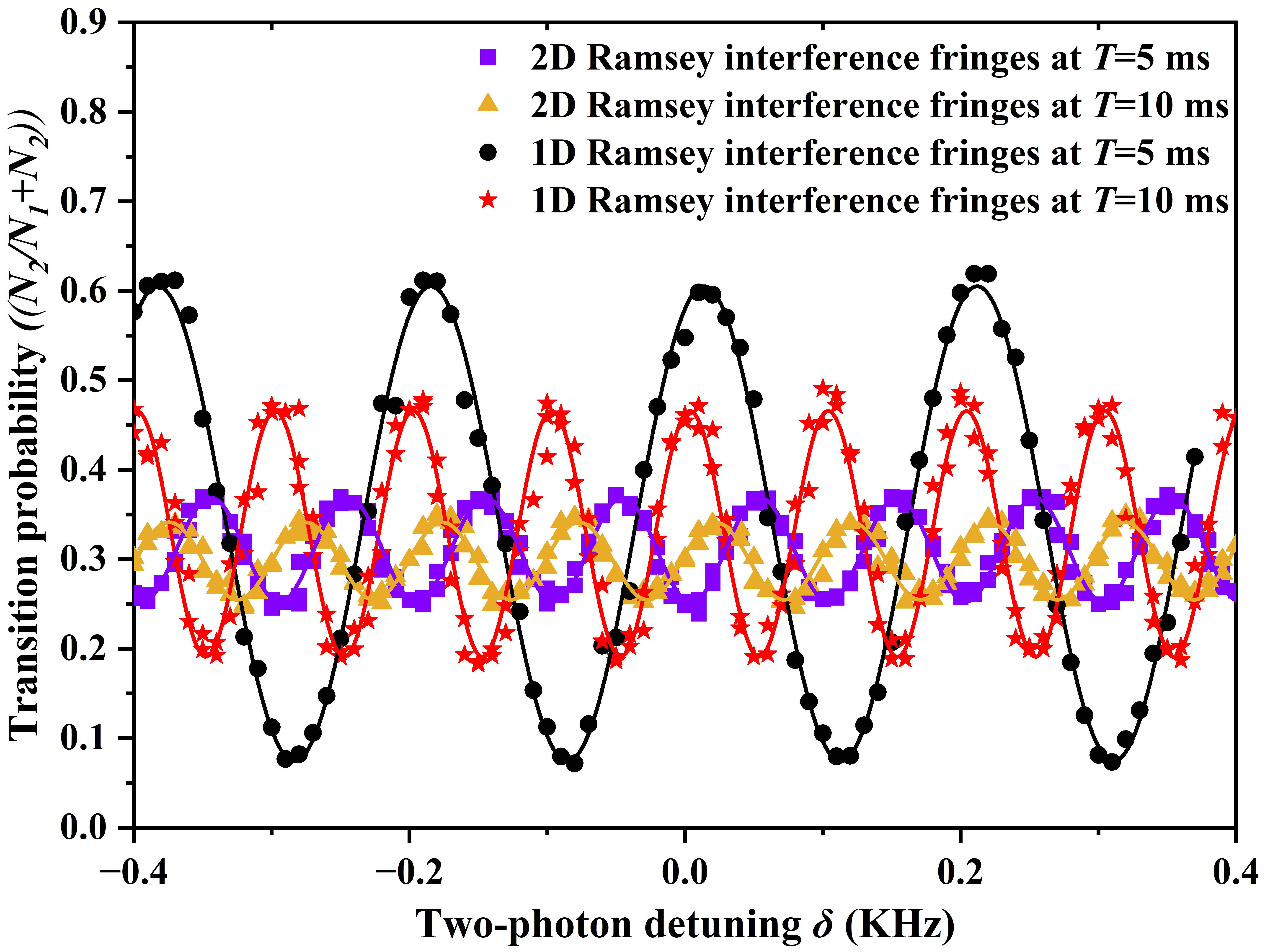}
	\caption{Ramsey interference fringes measured in the single-dimensional and two-dimensional Raman configurations with different pulse separation times $T$. The purple squares and fitted curve correspond to the two-dimensional Ramsey fringes at $T$=5 ms, the black circles and fitted curve to the one-dimensional Ramsey fringes at $T$=5 ms, the red pentagrams and fitted curve to the one-dimensional Ramsey fringes at $T$=10 ms, and the yellow triangles and fitted curve to the two-dimensional Ramsey fringes at $T$=10 ms.
	}
	\label{Ramsey}
\end{figure}

The single-dimensional and two-dimensional Ramsey-Bordé interference fringes are shown in Fig.~\ref{Ramsey} for $T=5~\mathrm{ms}$ and $T=10~\mathrm{ms}$. In the two-dimensional configuration, the measured contrasts are $23.2\%$ and $14.8\%$ (purple square markers and fitted curve for $T=5~\mathrm{ms}$; yellow triangle markers and fitted curve for $T=10~\mathrm{ms}$), compared with $53\%$ and $27\%$ in the single-dimensional configuration (black circle markers and fitted curve for $T=5~\mathrm{ms}$; red pentagram markers and fitted curve for $T=10~\mathrm{ms}$). The two-dimensional contrast is therefore reduced to about $44\%$--$55\%$ of the corresponding single-dimensional value. Additional measurements at $T=3~\mathrm{ms}$ and $T=4~\mathrm{ms}$ give two-dimensional contrasts of $28.8\%$ and $25.2\%$, while the single-dimensional contrasts remain approximately $57\%$. The contrast decreases with increasing pulse separation time in both configurations. The atomic velocity in the Raman interaction region is approximately 1.07 m/s. The diameter of Raman beams is 18 mm, while the initial atomic cloud radius is about 1 mm and expands to approximately 3.44 mm at the time of interrogation. In addition, the Raman intensity follows a Gaussian spatial profile, resulting in reduced transition efficiency for atoms near the beam edges. Taking this effect into account, the diameter of Raman beams  is estimated to be approximately 11 mm. Under these conditions, an interrogation time of $T$ = 10 ms approaches the maximum value permitted by the finite Raman beams size. The subsequent reduction in fringe contrast is therefore attributed primarily to the limited spatial extent of the Raman beams.

Following the observation of two-dimensional Ramsey-Bordé interference, the Raman configuration was used to explore a two-dimensional three-pulse M-Z sequence (as shown in Fig.~\ref{experimental apparatus}(b)). In this configuration, reliable M-Z interference fringes could not be extracted. This result can be understood from the Raman transition measurements discussed above. As shown in Fig.~\ref{Raman spectrum and Raman Rabi}(d), even for the co-propagating Raman transition, where the accessible momentum-state structure is relatively simple, the maximum transition probability at the nominal single-dimensional $\pi$-pulse position in the two-dimensional configuration is only about $53\%$ of that in the corresponding single-dimensional case. This value provides a rough estimate of the fraction of atoms that can participate in the desired momentum-state pathways in a three-pulse M-Z sequence, without including additional population transfer into cross-dimensional or reverse momentum states. Therefore, the contrast of the two-dimensional M-Z interferometer would be reduced by at least a factor of $(0.53)^3\approx0.15$ compared with the single-dimensional case.

The main limitation is the action of the second $\pi$ pulse in the two-dimensional Raman beams. As shown in Fig.~\ref{principles}(c), in addition to redirecting the target interferometer arms, this pulse drives transitions among the $\ket{F=2}$ momentum states. Consequently, a large fraction of atoms is redistributed among multiple $\ket{F=2}$ momentum states, which cannot be distinguished by conventional state-selective detection. The detected population thus contains contributions from the target interferometer as well as from parasitic momentum pathways within the $\ket{F=2}$ manifold, strongly suppressing the observable fringe contrast. This explains why stable two-dimensional Mach-Zehnder fringes could not be obtained under the present experimental conditions.
\subsection{Simulation results}
\begin{figure}[ht]
	\centering\includegraphics[width=9cm]{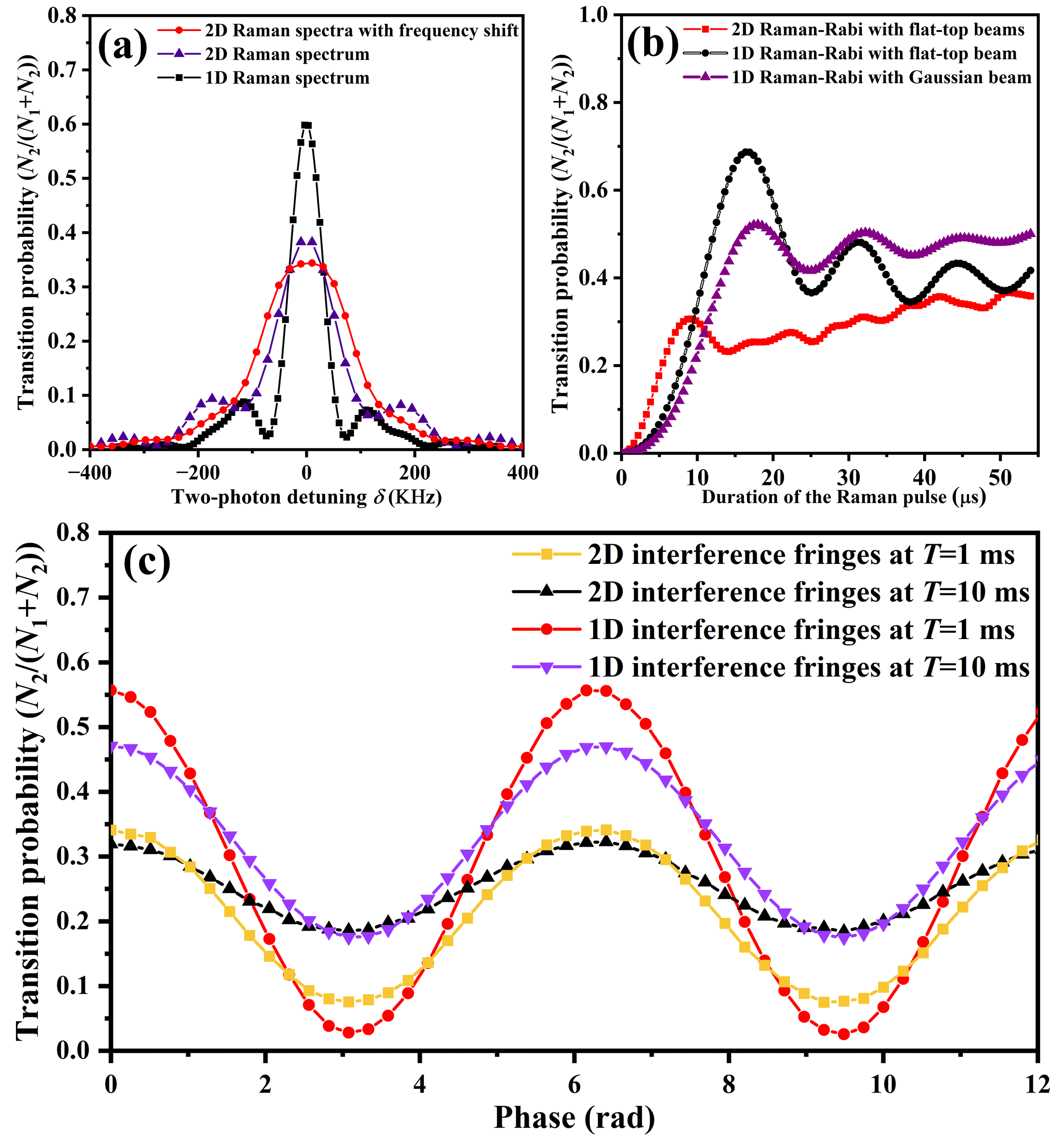}
	\caption{Simulation results for atomic dynamics driven by multidimensional Raman fields.
(a) Co-propagating Raman spectra. The black dotted line denotes the single-dimensional Raman spectrum, the purple dotted line represents the simulated two-dimensional Raman transition without considering the AC Stark shift, and the red dotted line represents the simulated two-dimensional Raman transition including the AC Stark shift induced by the intensity imbalance between the two dimensions; (b) co-propagating Raman Rabi oscillations. The red dotted line represents the two-dimensional Raman Rabi oscillation with flat-top beams, while the black and purple dotted lines represent the single-dimensional Raman Rabi oscillation with flat-top and Gaussian beams; (c) Simulated Ramsey fringes for different Raman pulse separation times \(T\). The red and yellow dotted lines represent the single and two-dimensional Ramsey fringes at \(T=1~\mathrm{ms}\), respectively, while the purple and black dotted lines represent the single and two-dimensional Ramsey fringes at \(T=10~\mathrm{ms}\), respectively.}
	\label{simulation result}
\end{figure}

To validate the multidimensional Raman interferometry theory, we implemented a numerical simulation framework based on the multidimensional atom-optics formalism described above and in the Supplemental Document. In particular, we performed detailed simulations of the co-propagating Raman interaction and the corresponding interferometric process, in order to gain a deeper understanding of the underlying mechanism of multidimensional atom interferometry. The calculated Raman spectra, Rabi oscillations, and Ramsey interference fringes are compared with the corresponding experimental measurements in Fig.~\ref{simulation result}.

In the simulation, the initial atomic ensemble parameters were chosen to match the experimental conditions. The atomic temperature was set to $4~\upmu\mathrm{K}$, and the initial atomic velocity distribution was generated according to the experimentally measured launch and cooling conditions. The single-photon detunings of the Raman beams along both dimensions were set to $400~\mathrm{MHz}$. The two Raman dimensions were arranged at an intersection angle of $36^\circ$, and the bias magnetic field was oriented along the bisector of the two Raman wave vectors to define the quantization axis. The Raman beams had approximate intensities of $150~\mathrm{mW}$ each, with an intensity ratio of $I_2/I_1 \simeq 1.8$. The Raman pulse separation time was $T=10~\mathrm{ms}$.

As shown in Fig.~\ref{simulation result}(a), we simulated the co-propagating Raman spectra for both single and two-dimensional configurations. Based on the previously introduced assumption of optical power attenuation caused by multiple reflections, an additional AC Stark shift of $\pm50~\mathrm{kHz}$ was introduced into the cross-dimensional Raman interaction Hamiltonian to account for the laser power loss in the experimental setup. The black and red curves show good agreement with the experimental spectra presented in Fig.~\ref{Raman spectrum and Raman Rabi}(a), demonstrating that the model captures the dominant physical mechanisms of the multidimensional Raman interaction.
For comparison, the purple curve shows the simulated spectrum without including the AC Stark shift, assuming ideal Gaussian Raman beams. In this case, the Raman spectrum exhibits the characteristic sinc-function side lobes in the frequency domain, which arise from the Fourier transform of a Gaussian-shaped temporal pulse~\cite{butts2013efficient}.

Fig.~\ref{simulation result}(b) presents the simulated Raman Rabi oscillations for the single and two-dimensional configurations. The simulations use a top-hat beam approximation, since the central region of the Raman beams, with a waist of approximately $15~\mathrm{mm}$, can be reasonably approximated as a uniform intensity profile. In the simulation, the black dotted line represents the single-dimensional Raman Rabi oscillation with a top-hat beam, the red dotted line represents the two-dimensional Raman Rabi oscillation with top-hat beams for both dimensions, and the purple dotted line represents the single-dimensional Raman Rabi oscillation with a Gaussian beam.

In the single-dimensional configuration (black curve), a clear coherent Rabi oscillation is observed, with the first $\pi$-pulse occurring at approximately $13~\upmu\mathrm{s}$. In contrast, in the two-dimensional configuration (red curve), the Rabi oscillation amplitude is significantly lower from the outset, and the transition probability quickly reaches an inflection point around $8~\upmu\mathrm{s}$. No clear subsequent oscillation is observed, indicating that in the multidimensional Raman field, atoms are partially transferred into additional momentum states through cross-dimensional Raman coupling channels, which reduces the coherent population exchange between the target states. The resulting simulations are in good agreement with the experimental observations in Fig.~\ref{Raman spectrum and Raman Rabi}(c).

As shown in Fig.~\ref{simulation result}(c), we simulated the Raman Ramsey interference fringes for both the single- and two-dimensional configurations at pulse separation times of $T=1~\mathrm{ms}$ and $T=10~\mathrm{ms}$. The simulations reproduce the reduction of Ramsey fringe contrasts in the two-dimensional configuration. For $T=1~\mathrm{ms}$, the simulated contrasts are approximately $60\%$ and $27\%$ for the single and two-dimensional cases, respectively. When the pulse separation time is increased to $T=10~\mathrm{ms}$, the corresponding contrasts decrease to about $29\%$ and $14\%$. These results are in reasonable agreement with the experimental observations. The residual discrepancy between the simulated and measured contrasts is mainly attributed to the difference between the idealized beam model used in the simulation and the actual Raman beams profile in the experiment. In the simulation, the flat-top Raman beams were approximated by a Gaussian beam with a waist radius of $24~\mathrm{mm}$, truncated by an aperture with a diameter of $15~\mathrm{mm}$. Therefore, the simulated contrast is sensitive to the assumed beam profile, aperture size, and the exact timing at which the atoms interact with the Raman beams.

\section{Discussion}
\begin{figure}[ht]
	\centering\includegraphics[width=8cm]{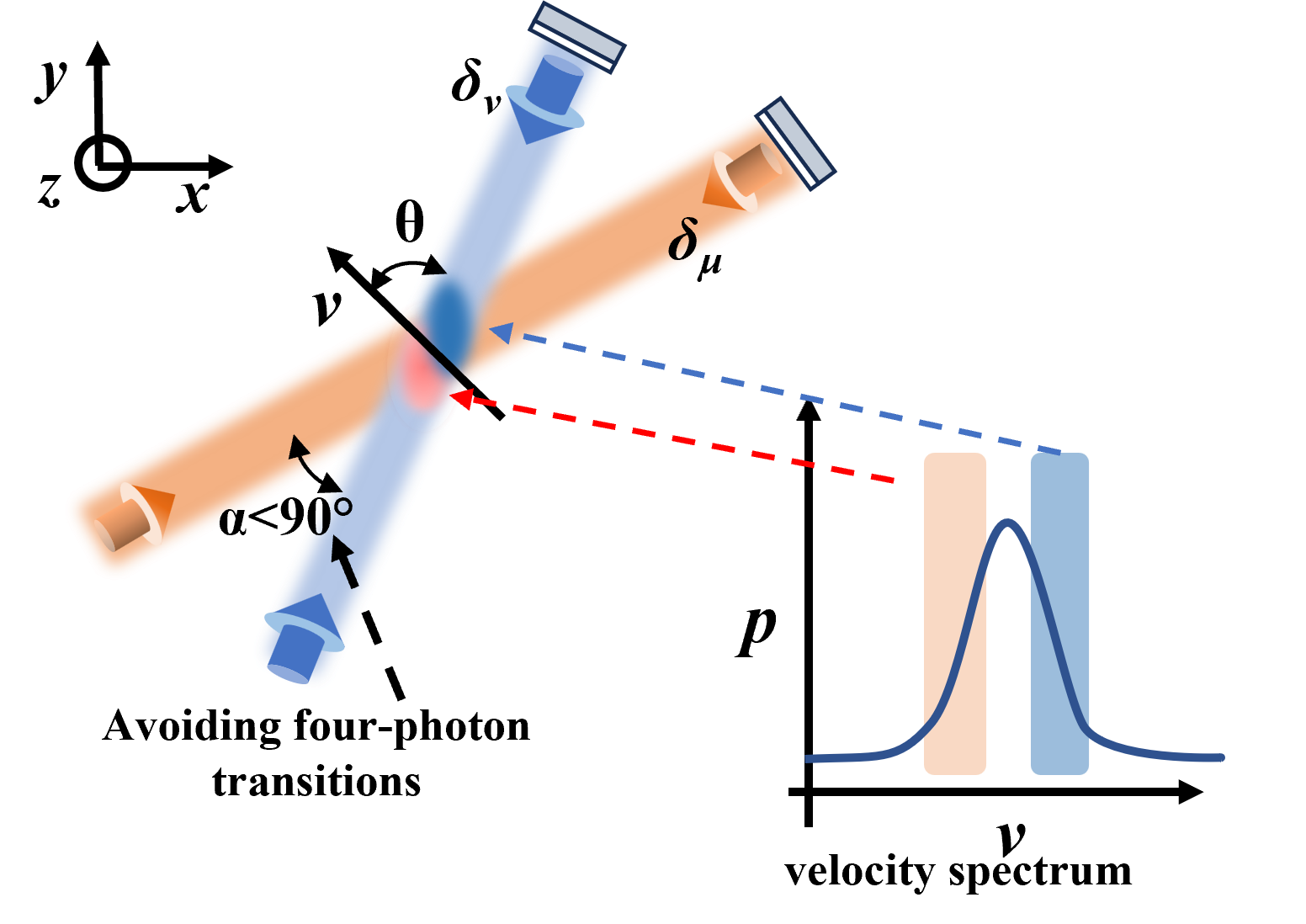}
	\caption{The schematic of two-dimensional atomic interferometer.}
	\label{discuss}
\end{figure}
The reduction of undesired Raman transitions is a central requirement for two-dimensional Raman atom interferometry. In a multidimensional Raman field, two different mechanisms should be distinguished. The first mechanism is the cross-dimensional two-photon Raman coupling formed directly by Raman field components belonging to different dimensions. For example, the Raman field component acting on the $|a\rangle$ side in dimension $i$ and the Raman field component acting on the $|b\rangle$ side in dimension $j$ can jointly form an effective Raman coupling $V_{ij}$. This process is not part of the intended single-dimensional Raman coupling, but arises from the simultaneous presence of multiple Raman field components. Its effective Rabi frequency is determined by the product of the two single-photon Rabi frequencies and the corresponding single-photon detunings. Therefore, within the reduced three-level approximation, one possible way to suppress the cross-dimensional Raman coupling is to make the two virtual-transition contributions in the effective coupling coefficient cancel each other. As indicated by Eq.~\ref{15}, when the two single-photon detunings approximately satisfy $\Delta^i_{a}=-\Delta^j_{b}$,
the cross-dimensional Raman coupling $V_{ij}$ is strongly suppressed. This condition should be regarded as a first-order suppression criterion in the simplified model. However, the cancellation condition can be modified by the excited-state hyperfine structure, polarization-dependent coupling strengths, and Zeeman shifts. Therefore, this condition should be regarded as a first-order design rule and should be evaluated using the full coupling matrix in the numerical model.

Sequential reverse Raman transitions can also reduce the contrast of the intended interferometer. For example, atoms transferred by one Raman dimension,
$|F=1,\mathbf p\rangle\rightarrow |F=2,\mathbf p+\hbar\boldsymbol{\kappa}_{\mu}\rangle$,
may subsequently undergo a reverse Raman transition driven by another dimension,
$|F=2,\mathbf p+\hbar\boldsymbol{\kappa}_{\mu}\rangle\rightarrow
|F=1,\mathbf p+\hbar\boldsymbol{\kappa}_{\mu}-\hbar\boldsymbol{\kappa}_{\nu}\rangle$.
This mechanism is different from the direct cross-dimensional two-photon coupling discussed above, because it occurs after atoms have already been transferred by one Raman dimension. It transfers atoms from the desired output momentum state to an undesired momentum state, thereby reducing the population participating in the intended interference and lowering the fringe contrast.

This sequential process can be suppressed by separating the Raman resonance conditions of the two dimensions in velocity space. When the two Raman beams geometries are arranged with a finite angle, the Doppler shift of each Raman transition depends on the projection of the atomic velocity onto the corresponding effective wave vector (As shown in Fig.~\ref{discuss}). The atomic ensemble has a single velocity distribution, and both Raman dimensions act on this same ensemble. The key idea is not that the two dimensions naturally produce different Doppler detunings, but that their two-photon detunings are deliberately chosen to select different velocity classes within the same distribution. In this way, one Raman dimension is resonant with atoms on one side of the velocity distribution, while the other dimension is resonant with atoms on another side. When $\delta_{\nu}\neq\delta_{\mu}$ ($\delta_{\nu}$ and $\delta_{\mu}$ are the corresponding velocity-dependent detunings), the two Raman dimensions address different portions of the same velocity distribution. This velocity-class separation reduces the probability that atoms transferred by one Raman dimension remain resonant with the reverse Raman transition driven by the other dimension. As a result, the undesired inter-dimensional population transfer is suppressed, and the mutual influence between the two Raman dimensions is reduced. The benefit is that the reverse transition between dimensions is suppressed. However, this velocity-selective scheme reduces the number of atoms participating in each interferometer, because only a selected portion of the velocity distribution is addressed. This limitation could be further alleviated by programmable multi-frequency Raman fields, which enable spectral-domain coherent control of broadband Raman coupling and allow a broader velocity distribution to participate in the interferometer~\cite{wang2026spectral}. This provides a possible route to improve the signal-to-noise ratio while maintaining the velocity-class separation required to suppress inter-dimensional transitions.

A further way to separate the two dimensions is to use internal-state selectivity. By preparing atoms in different magnetic sublevels, the two Raman dimensions can be assigned to different Zeeman-sensitive transitions. For example, atoms may be prepared in $|F=1,m_F=0\rangle$ and then transferred to $|F=2,m_F=\pm1\rangle$ using radio-frequency or microwave coupling. One Raman dimension can then address the $|m_F=1\rangle$ state, while the other addresses the $|m_F=-1\rangle$ state. The bias magnetic field shifts the two Raman resonances differently, providing a spectral separation between the two dimensions. This can further reduce the probability that atoms transferred by one dimension are resonantly driven back by the other dimension.

However, the Zeeman-based separation also introduces magnetic-field sensitivity. The first-order Zeeman shift produces a resonance-frequency bias, which must be calibrated or canceled by differential measurement. Therefore, this method is useful when a stable quantization field is available and when the residual Zeeman-induced phase shift can be treated as a common-mode or calibratable contribution.

\section{Conclusion}

In summary, this work presents a systematic theoretical, numerical, and experimental investigation of multidimensional atom interferometry from the perspective of atom-laser interactions. Building upon a full physical-process simulation platform for cold atom interferometry developed previously, we model multidimensional Raman transitions and momentum-state transfer processes, and construct an effective Hamiltonian framework in the multidimensional momentum-state basis. This framework provides a unified description of intended Raman transitions, cross-dimensional coupling channels, AC Stark shifts, Doppler shifts, recoil effects, and higher-order momentum pathways, revealing that multidimensional Raman atom interferometry cannot be treated as a simple superposition of independent one-dimensional interferometers.

Experimentally, a self-developed fountain-type cold atom interferometer is used to investigate both co-propagating and counter-propagating Raman transitions in one-dimensional and two-dimensional configurations. The measured Raman spectra demonstrate that the two-dimensional Raman configuration leads to spectral broadening and reduced transition efficiency compared with the one-dimensional case. Rabi oscillation measurements indicate that the coherent population exchange between target momentum states is suppressed in the two-dimensional configuration, consistent with the presence of reverse Raman transitions and population redistribution into additional momentum states. Through comparative measurements, two-dimensional Ramsey interference fringes are successfully observed, with a fringe contrast of $15.8\%$ at an interrogation time of $T=10~\mathrm{ms}$.

Numerical simulations based on the effective Hamiltonian reproduce the main experimental observations. The simulated Raman spectra and Rabi oscillations show good agreement with the measurements, capturing both spectral broadening and reduced Rabi contrast. For the Ramsey interferometer with $T=10~\mathrm{ms}$, the simulation gives fringe contrasts of approximately $14\%$ and $29\%$ for the two-dimensional and one-dimensional configurations, respectively. While the exact optical power distribution in the experiment is not fully known and may lead to quantitative differences, the overall agreement supports the validity of the multidimensional Hamiltonian model and clarifies the dominant mechanisms limiting interference contrast.

These results indicate that the dominant limitations to transition efficiency in multidimensional Raman atom interferometry arise from reverse Raman transitions between different dimensions and cross-dimensional effective Raman couplings. The former can be mitigated via velocity-space separation, Zeeman-state selectivity, or optimized detuning strategies, while the latter requires careful suppression to avoid parasitic momentum transfer. Together with the theoretical analysis, numerical simulations, and experimental demonstrations, this work provides a comprehensive framework for multidimensional atom interferometry and may facilitate the development of high-efficiency quantum inertial sensors, including atom-interferometric full-tensor gravity gradiometers. The simulation code developed in this work has been organized and released as an open-source package~\cite{zhao2026multidimensional}.

\begin{backmatter}
\bmsection{Funding}
This work was supported by the National Key Research and Development Program of China under Grant 2024YFB3908102 and Grant 2023YFC2907000.

\bmsection{Disclosures}
The authors declare no conflicts of interest.

\bmsection{Data Availability Statement}
Data underlying the results presented in this paper are not publicly available at this time but may be obtained from the authors upon reasonable request.
\end{backmatter}


\bibliography{sample}

@inproceedings{mukherjee2019conceptualization,
  title={Conceptualization, design and prospects of atom-interferometry based full tensor gravity gradiometer},
  author={Mukherjee, Anirban and Rakonjac, Ana and Kumar, Ravi},
  booktitle={Frontiers in Optics},
  pages={FS3A--4},
  year={2019},
  organization={Optica Publishing Group}
}

@article{kasevich1991atomic,
  title={Atomic interferometry using stimulated Raman transitions},
  author={Kasevich, Mark and Chu, Steven},
  journal={Physical review letters},
  volume={67},
  number={2},
  pages={181},
  year={1991},
  publisher={APS}
}

@article{borde1989atomic,
  title={Atomic interferometry with internal state labelling},
  author={Bord{\'e}, Ch J},
  journal={Physics letters A},
  volume={140},
  number={1-2},
  pages={10--12},
  year={1989},
  publisher={Elsevier}
}

@misc{zhao2026multidimensional,
  author       = {Zhao, Yingpeng},
  title        = {Multidimensional atomic interference program},
  year         = {2026},
  howpublished = {figshare. Software},
  doi          = {10.6084/m9.figshare.32215569.v1},
  url          = {https://doi.org/10.6084/m9.figshare.32215569.v1}
}

@article{bongs2019taking,
  title={Taking atom interferometric quantum sensors from the laboratory to real-world applications},
  author={Bongs, Kai and Holynski, Michael and Vovrosh, Jamie and Bouyer, Philippe and Condon, Gabriel and Rasel, Ernst and Schubert, Christian and Schleich, Wolfgang P and Roura, Albert},
  journal={Nature Reviews Physics},
  volume={1},
  number={12},
  pages={731--739},
  year={2019},
  publisher={Nature Publishing Group UK London}
}

@article{butts2013efficient,
  title={Efficient broadband Raman pulses for large-area atom interferometry},
  author={Butts, David L and Kotru, Krish and Kinast, Joseph M and Radojevic, Antonije M and Timmons, Brian P and Stoner, Richard E},
  journal={Journal of the Optical Society of America B Optical Physics},
  volume={30},
  number={4},
  pages={922},
  year={2013}
}

@article{geiger2020high,
  title={High-accuracy inertial measurements with cold-atom sensors},
  author={Geiger, Remi and Landragin, Arnaud and Merlet, S{\'e}bastien and Pereira Dos Santos, Franck},
  journal={AVS Quantum Science},
  volume={2},
  number={2},
  year={2020},
  publisher={AIP Publishing}
}

@article{tino2021testing,
  title={Testing gravity with cold atom interferometry: results and prospects},
  author={Tino, Guglielmo M},
  journal={Quantum Science \& Technology},
  volume={6},
  number={2},
  pages={024014},
  year={2021},
  publisher={IOP Publishing}
}

@article{cronin2009optics,
  title={Optics and interferometry with atoms and molecules},
  author={Cronin, Alexander D and Schmiedmayer, J{\"o}rg and Pritchard, David E},
  journal={Reviews of Modern Physics},
  volume={81},
  number={3},
  pages={1051--1129},
  year={2009},
  publisher={APS}
}

@article{rakholia2014dual,
  title={Dual-axis high-data-rate atom interferometer via cold ensemble exchange},
  author={Rakholia, Akash V and McGuinness, Hayden J and Biedermann, Grant W},
  journal={Physical Review Applied},
  volume={2},
  number={5},
  pages={054012},
  year={2014},
  publisher={APS}
}

@article{wu2017multiaxis,
  title={Multiaxis atom interferometry with a single-diode laser and a pyramidal magneto-optical trap},
  author={Wu, Xuejian and Zi, Fei and Dudley, Jordan and Bilotta, Ryan J and Canoza, Philip and M{\"u}ller, Holger},
  journal={Optica},
  volume={4},
  number={12},
  pages={1545--1551},
  year={2017},
  publisher={Optical Society of America}
}

@article{salducci2024quantum,
  title={Quantum sensing of acceleration and rotation by interfering magnetically launched atoms},
  author={Salducci, Cl{\'e}ment and Bidel, Yannick and Cadoret, Malo and Darmon, Sarah and Zahzam, Nassim and Bonnin, Alexis and Schwartz, Sylvain and Blanchard, C{\'e}dric and Bresson, Alexandre},
  journal={Science Advances},
  volume={10},
  number={44},
  pages={eadq4498},
  year={2024},
  publisher={American Association for the Advancement of Science}
}

@article{stolzenberg2025multi,
  title={Multi-axis inertial sensing with 2d matter-wave arrays},
  author={Stolzenberg, K and Struckmann, C and Bode, S and Li, R and Herbst, A and Vollenkemper, V and Thomas, D and Rajagopalan, A and Rasel, EM and Gaaloul, N and others},
  journal={Physical Review Letters},
  volume={134},
  number={14},
  pages={143601},
  year={2025},
  publisher={APS}
}

@article{barrett2019multidimensional,
  title={Multidimensional atom optics and interferometry},
  author={Barrett, Brynle and Cheiney, Pierrick and Battelier, Baptiste and Napolitano, Fabien and Bouyer, Philippe},
  journal={Physical Review Letters},
  volume={122},
  number={4},
  pages={043604},
  year={2019},
  publisher={APS}
}

@article{zhao2025construction,
  title={Construction of a program for physical simulation of cold atom interferometry},
  author={Zhao, Yingpeng and Zhang, Cheng and Bao, Shuning and Li, Dianrong and Niu, Jingyu and Tong, Wenjian and Cheng, Bing and Wang, Xiaolong and Weng, Kanxing and Li, Hao and others},
  journal={Optics Express},
  volume={33},
  number={3},
  pages={6373--6391},
  year={2025},
  publisher={Optica Publishing Group}
}

@article{zhao2024implementation,
  title={The implementation of a compact cold atom interference gyroscope based on miniaturized quartz vacuum chamber},
  author={Zhao, Yingpeng and Li, Dianrong and Niu, Jingyu and Bao, Shuning and Zhang, Kaijun and Wang, Yuchen and Cheng, Bing and Zhang, Cheng and Niu, Kexiao and Liu, Yuanzheng and others},
  journal={IEEE Sensors Journal},
  volume={24},
  number={24},
  pages={40507--40517},
  year={2024},
  publisher={IEEE}
}

@article{fixler2007atom,
  title={Atom interferometer measurement of the Newtonian constant of gravity},
  author={Fixler, Jeffrey B and Foster, GT and McGuirk, JM and Kasevich, MA},
  journal={Science},
  volume={315},
  number={5808},
  pages={74--77},
  year={2007},
  publisher={American Association for the Advancement of Science}
}

@article{prevedelli2014measuring,
  title={Measuring the Newtonian constant of gravitation G with an atomic interferometer},
  author={Prevedelli, Marco and Cacciapuoti, L and Rosi, G and Sorrentino, F and Tino, GM},
  journal={Philosophical Transactions of the Royal Society A: Mathematical, Physical and Engineering Sciences},
  volume={372},
  number={2026},
  year={2014},
  publisher={The Royal Society}
}

@article{shettell2024geophysical,
  title={Geophysical survey based on hybrid gravimetry using relative measurements and an atomic gravimeter as an absolute reference},
  author={Shettell, Nathan and Lee, Kai Sheng and Oon, Fong En and Maksimova, Elizaveta and Hufnagel, Christoph and Wei, Shengji and Dumke, Rainer},
  journal={Scientific Reports},
  volume={14},
  number={1},
  pages={6511},
  year={2024},
  publisher={Nature Publishing Group UK London}
}

@inproceedings{jia2025closed,
  title={Closed-Loop Atomic Interferometric Inertial Sensor for Dynamic Measurement},
  author={Jia, Weichen and Yan, Peiqiang and Shen, Ascar and Feng, Yanying},
  booktitle={2025 IEEE International Symposium on Inertial Sensors and Systems (INERTIAL)},
  pages={1--4},
  year={2025},
  organization={IEEE}
}

@article{stray2022quantum,
  title={Quantum sensing for gravity cartography},
  author={Stray, Ben and Lamb, Andrew and Kaushik, Aisha and Vovrosh, Jamie and Rodgers, Anthony and Winch, Jonathan and Hayati, Farzad and Boddice, Daniel and Stabrawa, Artur and Niggebaum, Alexander and others},
  journal={Nature},
  volume={602},
  number={7898},
  pages={590--594},
  year={2022},
  publisher={Nature Publishing Group UK London}
}

@article{peters2001high,
  title={High-precision gravity measurements using atom interferometry},
  author={Peters, Achim and Chung, Keng Yeow and Chu, Steven},
  journal={Metrologia},
  volume={38},
  number={1},
  pages={25--61},
  year={2001}
}

@article{liu2024multidimensional,
  title={Multidimensional matter-wave beam splitters by multiphoton hyperfine Raman transitions},
  author={Liu, Xu-Dong and Chen, Zong-You and Huang, Wei-Chun and Lee, Yang-Kao and Chen, Yong-Fan and Kuan, Pei-Chen},
  journal={Physical Review A},
  volume={109},
  number={1},
  pages={013327},
  year={2024},
  publisher={APS}
}

@article{wu2024construction,
  title={Construction of absolute gravity benchmark offshore with an atomic gravimeter},
  author={Wu, Bin and Zhao, Yingpeng and Zhou, Yin and Yuan, Wenwen and Li, Dianrong and Bao, Shuning and Zhu, Dong and Cheng, Bing and Wu, Leyuan and Zhou, Jiangcun and others},
  journal={IEEE Sensors Journal},
  volume={24},
  number={15},
  pages={23527--23536},
  year={2024},
  publisher={IEEE}
}

@article{chang2023multiphoton,
  title={Multiphoton hyperfine raman transitions with different quantization axes},
  author={Chang, Jia-Wei and Huang, Yu-Chieh and Xu, Zhen-Wei and Kan, Yuan-Hung and Huang, Hsing-Han and Huang, Jia-Zhun and Fu, Kai-Jun and Liu, Xu-Dong and Kuan, Pei-Chen},
  journal={Physical Review A},
  volume={108},
  number={1},
  pages={013308},
  year={2023},
  publisher={APS}
}

@article{kasevich1992measurement,
  title={Measurement of the gravitational acceleration of an atom with a light-pulse atom interferometer},
  author={Kasevich, Mark and Chu, Steven},
  journal={Applied Physics B},
  volume={54},
  number={5},
  pages={321--332},
  year={1992},
  publisher={Springer}
}

@article{shao2025performance,
  title={Performance Evaluation of a Compact Cold Atom Interferometer Gyroscope},
  author={Shao, Yiyang and Zhang, Yuanlong and Tong, Wenjian and Gao, Haitao and Bao, Shuning and Zhang, Cheng and Niu, Jingyu and Li, Dianrong and Zhao, Yingpeng and Yin, Ruiduo and others},
  journal={IEEE Sensors Journal},
  year={2025},
  publisher={IEEE}
}

@article{wang2026spectral,
  title={Spectral-Domain Coherent Control of Broadband Raman Coupling in Atom Interferometry},
  author={Wang, Sheng-Zhe and Jia, Wei-Chen and Xin, Yue and Cai, Qian-Lan and Zhao, Yingpeng and Feng, Yan-Ying},
  journal={arXiv preprint arXiv:2604.03629},
  year={2026}
}

\end{document}